\documentclass[amssymb,superscriptaddress,eqsecnum,11pt,secnumarabic]{revtex4}

\renewcommand{\epsilon}{\varepsilon}
\def\bi{\mathbb{I}}
\def\Z{\mathbb{Z}}
\def\pf{{\rm Pf\ }}
\def\la{\langle}
\def\ra{\rangle}
\def\un#1{{\underline{#1}}}

\def\be{\begin{equation}}
\def\ee{\end{equation}}
\def\bea{\begin{eqnarray}}
\def\eea{\end{eqnarray}}
\newcommand\egal{&\!\!=\!\!&}

\def\cell{\begin{picture}(25,25)(-5,3)
\put(0,0){\line(1,0){15}}
\put(0,0){\line(0,1){15}}
\put(0,15){\line(1,0){15}}
\put(15,0){\line(0,1){15}}
\put(0,-4){\makebox(0,0)[c]{\footnotesize $i_4$}}
\put(15,-4){\makebox(0,0)[c]{\footnotesize $i_3$}}
\put(0,20){\makebox(0,0)[c]{\footnotesize $i_1$}}
\put(15,20){\makebox(0,0)[c]{\footnotesize $i_2$}}
\end{picture}}

\begin{document}

\title{\Large Boundary monomers in the dimer model}
\author{Vyatcheslav B. Priezzhev}
\affiliation{Bogolubov Laboratory of
Theoretical Physics, Joint Institute for Nuclear Research, 141980
Dubna, Russia}
\author{Philippe Ruelle}
\affiliation{Institut de Physique Th\'eorique,
Universit\'e catholique de Louvain,\\ 1348 Louvain-La-Neuve, Belgium
\vspace{.5cm}
}

\begin{abstract}
The correlation functions of an arbitrary number of boundary monomers in the system of close-packed dimers on the square lattice are computed exactly in the scaling limit. The equivalence of the $2n$-point correlation functions with those of a complex free fermion is proved, thereby reinforcing the description of the monomer-dimer model by a conformal free field theory with central charge $c=1$.
\end{abstract}


\maketitle

\section{Introduction}

The dimer model has been originally introduced to describe physical adsorption of
diatomic molecules on crystal surfaces \cite{foru}. The first studies of the dimer
model are dated the early sixties, with pioneering works by Kasteleyn
\cite{kast}, Fisher \cite{fish}, Temperley and Fisher \cite{fite}, Ferdinand \cite{ferd} and Wu \cite{wu1}, who have studied the number of close-packed dimer configurations on specific (mostly square) finite bipartite lattices. Soon after that, correlations between dimers and monomers on the square lattice have been examined in \cite{fist}. The effects caused by the insertion of monomers on the square lattice have been reconsidered in a number of recent works \cite{tzwu,kong1,kong2,kong3}. Dimer and monomer correlations have also been revisited recently, with significantly different conclusions, on the triangular lattice (a non-bipartite graph) in \cite{fend}. 

A close-packed dimer configuration on a square grid is an arrangement of dimers such that a dimer covers two adjacent sites and every site is covered by exactly one dimer. As a generalization, one can look at the dimer configurations when some sites, called monomers, are not allowed to be covered by dimers. Corresponding to these arrangements of monomers and dimers, one considers the partition function 
\be
Z(w_h,w_v|z_1,\ldots,z_N) = \sum_{\rm coverings} w_h^{n_h} \, w_v^{n_v}.
\label{pf}
\ee
It counts the number of dimer coverings in presence of $N$ monomers located at
positions $z_1, \ldots,z_N$, in the bulk or on boundaries, with weights $w_h$ and $w_v$ assigned to horizontal and vertical dimers. As the number $n_h + n_v$ of dimers is fixed, the partition function essentially depends on $w_h,w_v$ through the ratio $w_h/w_v$ only. 

The dimer model belongs to the class of so-called free-fermion models \cite{FanWu}. It is well known that the partition function of the free-fermion models admits a representation in terms of fermionic Gaussian integrals which leads to determinant expressions for the partition and correlation functions. Despite the simple form of the lattice action in the free-fermion representation, the evaluation of correlation functions for certain physical observables can be more complicated because local variables correspond often to non-local fermion correlators. As illustrated in \cite{fist}, the monomer-monomer correlations in the dimer model need typically non-local computations if monomers are located in the bulk of the lattice. However the situation simplifies significantly for boundary monomers. In this article, we take advantage of this simplification and compute all monomer correlations exactly, in the scaling limit. We will show that in the continuum limit, the correlation functions for boundary monomers can be expressed in terms of complex free fermions located at sites occupied by the monomers. This implies that the boundary monomer degrees of freedom are described by a conformal field theory with central charge $c=1$. 

Our result complements previous results related to the description of the general dimer model by a conformal field theory. If a conformal field theory with central charge $c=-2$ accounts well for the dimer degrees of freedom \cite{iprh}, it seems that the full monomer-dimer model should correspond to a conformal theory with $c=1$. An early indication of this can be traced in the work of Au-Yang and Perk \cite{auperk}, who noted a close relationship between the 2-monomer correlator computed by Fisher and Stephenson \cite{fist} and the squared correlator of two Ising spins, equivalently the correlator in a doubled Ising model. Likewise dimer correlations on the square lattice have been more recently reinterpreted as correlators of two uncoupled massless Majorana fermions (and by massive Majorana fermions in the perturbation away from the square lattice to the triangular lattice) by Fendley et al \cite{fend}. More generally, Kenyon \cite{ken} has shown that a quantity associated with a dimer configuration, called the height function, converges in the scaling limit to a Gaussian free field (or free boson). Like the complex free fermion, the free boson corresponds to a conformal theory with central charge $c=1$.
The relation between the dimer model and free fermions has been revisited very recently by Dijkgraaf et al \cite{dijk}.


\section{General setting}

Here we briefly recall the way the partition function can be calculated when the graph is a finite portion of a square lattice. In the following, we set $w_h=w_v=1$ as we are not interested in the directional properties of the dimer coverings.

The simplest and best understood situation is when all sites of the domain ${\cal L}$ must covered by dimers, the so-called close-packed limit. In this case, the partition function, which simply counts the number of dimer configurations so that every site of ${\cal L}$ is covered by one dimer, can be expressed as the Pfaffian of an antisymmetric  matrix $K$,
\be
Z_{{\cal L}}^{(0)} = \pf K = +\sqrt{\det K}.
\ee
The sign is chosen so that the partition function is positive, and the matrix $K$ is a weighted adjacency matrix of ${\cal L}$. Each term in the expansion of the Pfaffian is naturally associated with a dimer covering, but the entries of $K$ have to be suitably chosen so that each covering contributes 1 to the partition function.

\begin{figure}
\setlength{\unitlength}{1mm}
\begin{picture}(100,50)(4,-5)
\multiput(0,0)(0,10){5}{\multiput(0,0)(10,0){10}{\vector(1,0){10}}
\put(100,0){\line(1,0){8}}}
\multiput(4,0)(20,0){6}{
\put(0,0){\vector(0,1){6}}
\multiput(0,6)(0,10){3}{\put(0,0){\vector(0,1){10}}}
\put(0,36){\line(0,1){8}}}
\multiput(14,0)(20,0){5}{\put(0,0){\line(0,1){5}}
\multiput(0,44)(0,-10){4}{\vector(0,-1){10}}}
\put(44,0){\circle*{2}}
\put(44,-4){\makebox(0,0)[c]{\small $x=0$}}
\put(114,0){\makebox(0,0)[c]{\small $y=1$}}
\end{picture}
\caption{Orientation of the bonds defining the signs in the Kasteleyn matrix $K$. }
\end{figure}
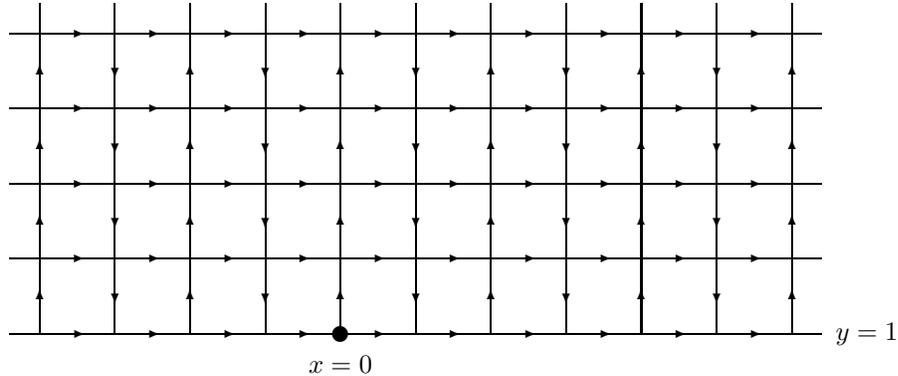

There are different choices for $K$, but a convenient one for what follows is the matrix originally considered by Kasteleyn. $K$ is an oriented adjacency matrix for ${\cal L}$, so that $K_{ij}=0$ if the sites $i,j$ are not nearest neighbours,
and $K_{ij}=\pm 1$ otherwise. The signs are given pictorially in Fig.1: an arrow
from $i$ to $j$ indicates that $K_{ij}=+1$ and $K_{ji}=-1$. The essential property of
$K$ is that the product of entries around any elementary cell of ${\cal L}$ is equal to $-1$,
\be
K_{i_1i_2} K_{i_2i_3} K_{i_3i_4} K_{i_4i_1} = -1, \qquad
{\rm for\ any\ cell\ }\cell.
\label{kast}
\ee
As $K$ is closely related to the Laplacian on ${\cal L}$, standard techniques like Fourier
series can be used to compute the determinant of $K$, and then its Pfaffian.

Monomers are sites which cannot be covered by dimers. Thus dimer configurations on ${\cal L}$
in presence of monomers at $z_1,z_2,\ldots,z_N$ are close-packed dimer coverings
of ${\cal L} \setminus \{z_1,z_2,\ldots\}$.

If all of the $N$ monomers are on the boundary of ${\cal L}$, the matrix $K(z_1,z_2,\ldots)$
defined from $K$ by removing the rows and columns labelled by the sites $z_1,z_2,\ldots$
is the Kasteleyn matrix for ${\cal L} \setminus \{z_1,z_2,\ldots\}$. As it still satisfies the
property (\ref{kast}), one has
\be
Z^{(N)}_{{\cal L}}(z_1,z_2,\ldots) = \pf K(z_1,z_2,\ldots) = +\sqrt{\det K(z_1,z_2,\ldots)}.
\ee

The monomer correlations are then defined by the ratios
\be
C(z_1,z_2,\ldots) \equiv {Z^{(N)}_{{\cal L}}(z_1,z_2,\ldots) \over Z^{(0)}_{{\cal L}}} = {\sqrt{\det K(z_1,z_2,\ldots)}
\over \sqrt{\det K}}.
\ee
We are eventually interested to compute these ratios in the thermodynamic limit.
It is then much more convenient to express $K(z_1,z_2,\ldots) = K + B(z_1,z_2,\ldots)$ as a finite rank perturbation of $K$, localized around the monomer positions. If indeed $B$ has rank $N$ (that is, $B_{ij}=0$ except if $i,j$ are in a set of $N$ sites), the ratio of infinite-dimensional determinants
\be
{Z^{(N)}_{{\cal L}}(z_1,z_2,\ldots) \over Z^{(0)}_{{\cal L}}} = \sqrt{\det K^{-1}K(z_1,z_2,\ldots)} =
\sqrt{\det(\bi + K^{-1}B)},
\label{correl}
\ee
reduces to a finite, rank $N$ determinant (which however involves entries of the infinite-dimensional matrix $K^{-1}$). 

The defect matrix $B$ has to satisfy two requirements: (i) if $k$ is a neighbouring site of
a monomer located at $z$, then $B_{z,k} = -K_{z,k} = -B_{k,z}$ so that $z$ is effectively
cut off from the rest of the grid; and (ii) the restriction of $B$ to the monomer sites
$z_1,z_2,\ldots$ must have a determinant equal to 1 (the simplest solution is to set $B_{z_i,z_i}=1$, but this is not always the most convenient way, see Section 4). All other entries of $B$ are equal to 0. The rank of $B$ increases linearly with the number of monomers, and so does the size of
the determinant.

When some of the monomers are away from the boundary, the situation changes dramatically.
Removing a non-boundary site from ${\cal L}$ creates a new elementary cell, around which the
product of the restricted $K$ matrix elements is not equal to $-1$. This can be remedied by changing the signs of $K$ along a path going from one monomer to another monomer.
This in effect introduces a non-local defect matrix, and complicates the calculation
since the size of the determinant increases with the distance between the monomers.
To date, the only known exact result on the square lattice is the old result by Fisher and Stephenson \cite{fist}, who proved that the 2-point correlation of bulk monomers in the scaling limit decays like $r^{-1/2}$. In contrast the monomer and dimer correlations decay exponentially on the triangular lattice \cite{fend}. There are indications that this last behaviour holds on non-bipartite graphs.


\section{Isolated monomers on a boundary}

Our purpose is to calculate the monomer correlations (\ref{correl}) for an arbitrary
number of isolated monomers on the boundary of the discrete upper half-plane (UHP), with no monomer away from the boundary. For definiteness, we take the boundary to be the
line $y=1$, so that the discrete UHP corresponds to $\{(x,y)\;:\; x \in \Z,\; y \in \Z_{>0}\}$. The positions of the monomers along the boundary are denoted by $(x_i,1)$, and we are interested in the scaling regime where all distances $x_{ij} \equiv x_i-x_j$ are large. We assume the $x_i$ are ordered from left to right, so that $x_{ij} < 0$ for $i<j$.

The first ingredient we need is the inverse of $K$. As mentioned in the previous section,
 the matrix $K$ itself is defined from Fig.1 where the rectangle is extended to the UHP.
The orientation of all horizontal bonds is to the right, while that of the vertical bonds
alternate; we fix the reference point by deciding that the vertical bonds on the line $x=0$
are oriented upwards. The matrix is then given by
\be
K_{(x,y),(x',y')} = [\delta_{x,x'-1} - \delta_{x,x'+1}] \delta_{y,y'} + (-1)^x \,
\delta_{x,x'} [\delta_{y,y'-1} - \delta_{y,y'+1}].
\ee

It is instructive and easy to compute the square of $K$,
\be
(-K^2)_{ij} = \cases{\hbox{number of nearest neighbours of $i, \quad$ for $i=j$},&\cr
\hbox{$-1, \quad$ if $i-j=\pm (2,0)$ or $\pm (0,2)$},&\cr
\hbox{$0 \quad $ otherwise.} &}
\ee
One sees that $-K^2$ connects sites within each of the four sublattices corresponding
to the parity of the $x$-coordinates and of the $y$-coordinates, and that its restriction
to any one of these is equal to a specific Laplacian. On the odd-odd and even-odd sublattices, the Laplacian is subjected to a closed boundary condition since the sites on the boundary ($y=1$) have three nearest neighbours. On the other two sublattices,
which have the line $y=2$ as boundary, along which the sites have four nearest neighbours,
the Laplacian is subjected to the open boundary condition. We can write
\be
-K^2 = \Delta^{\rm cl}_{\rm odd,odd} \oplus \Delta^{\rm cl}_{\rm even,odd} \oplus
\Delta^{\rm op}_{\rm odd,even} \oplus \Delta^{\rm op}_{\rm even,even}.
\ee

This allows to find the inverse of $K$,
\be
K^{-1} = - \Big[G^{\rm cl}_{\rm odd,odd} \oplus G^{\rm cl}_{\rm even,odd}
\oplus G^{\rm op}_{\rm odd,even} \oplus G^{\rm op}_{\rm even,even} \Big] \, K,
\ee
in terms of the well-known Green matrices $G = \Delta^{-1}$. The Green matrices
$G^{\rm cl}$ and $G^{\rm op}$ are related to the inverse Laplacian $G$ on the
full (discrete) plane by standard formulae,
\bea
G^{\rm cl}(x-x';y,y') &\equiv& G^{\rm cl}_{(x,y),(x',y')} = G(x-x',y-y') +
G(x-x',y+y'-1), \label{Gcl}\\
G^{\rm op}(x-x';y,y') &\equiv& G^{\rm cl}_{(x,y),(x',y')} = G(x-x',y-y') -
G(x-x',y+y').
\label{Gop}
\eea

One can now write the inverse of $K$ explicitly:
\bea
K^{-1}_{(x,y),(x',y')} &=& \Big[G^{\rm cl}({\textstyle{x-x'-1 \over 2};{y+1 \over 2},{y'+1
\over 2}}) - G^{\rm cl}({\textstyle{x-x'+1 \over 2};{y+1 \over 2},{y'+1 \over 2}})\Big],
\quad {\rm if\ } (x-x')y y'=1 \bmod 2, \nonumber\\
&& \hspace{-1cm} = \Big[G^{\rm op}({\textstyle{x-x'-1 \over 2};{y \over 2},{y' \over 2}}) -
G^{\rm op}({\textstyle{x-x'+1 \over 2};{y \over 2},{y' \over 2}})\Big],\quad {\rm if\ }
(x-x')(y-1)(y'-1)=1 \bmod 2, \nonumber\\
&& \hspace{-1cm} = (-1)^x \Big[G^{\rm cl}({\textstyle{x-x' \over 2};{y+1
\over 2},{y'+2 \over 2}}) - G^{\rm cl}({\textstyle{x-x' \over 2};{y+1 \over 2},{y'
\over 2}})\Big],\; {\rm if\ } (x-x'-1)y(y'-1)=1 \bmod 2, \nonumber\\
&& \hspace{-1cm} = (-1)^x \Big[G^{\rm op}({\textstyle{x-x' \over 2};{y \over 2},
{y'+1 \over 2}}) - G^{\rm op}({\textstyle{x-x' \over 2};{y \over 2},{y'-1 \over 2}})\Big],
\quad {\rm if\ } (x-x')(y-1)y'=1 \bmod 2. \nonumber\\
&& \hspace{-1cm} = 0, \quad \hbox{otherwise (when $x-x'$ and $y-y'$ have the same parity).}
\label{kinv}
\eea
By using the relations (\ref{Gcl}) and (\ref{Gop}), one may check that $K^{-1}$ is antisymmetric.

We are now ready to compute the relevant determinants (\ref{correl}). If there are $N$ monomers on the boundary, the $B$ matrix is $4N$-dimensional. A possible choice is to write it as the direct sum of $N$ 4-by-4 blocks $B_1(x_i)$, one for each monomer, so that
$B = \oplus_i B_1(x_i)$, with
\be
B_1(x) = \pmatrix{1 & 1 & -(-1)^x & -1 \cr -1 & 0 & 0 & 0 \cr
(-1)^x & 0 & 0 & 0 \cr 1 & 0 & 0 & 0},
\label{B1}
\ee
in the basis where the indices 1, 2, 3, 4 correspond respectively to the monomer itself,
its left, its upper and its right neighbour.

Using this explicit form of $B$ as well as the inverse of $K$, one may easily compute the
correlation functions, given by
\be
C(x_1,x_2,\ldots,x_N) \equiv {Z(x_1,x_2,\ldots,x_N) \over Z} = \sqrt{\det(\bi + K^{-1}B)}.
\ee
For large distances $|x_{ij}| \gg 1$, the asymptotic form of these correlators may be obtained by using the following expansions of the Green matrix on the plane,
\bea
G(m,k) \egal G(m,0) - {k^2 \over 4\pi m^2} + {k^4-3k^3 \over 8\pi m^4} + \ldots,\, \qquad k \ll m,\\
G(m,0) \egal -{1 \over 2\pi} \log{|m|} + \ldots
\label{expan}
\eea

We have computed explicitly the first few correlators, and found the following expressions. The 1- and 3-point functions are identically zero,
\be
C(x_1) = C(x_1,x_2,x_3) = 0,
\ee
as one would expect. On a finite rectangular grid (with an even number of sites), there must be an even number of monomers since otherwise, the rest of the rectangle cannot be covered with dimers.

The 2-point function is equal to, at dominant order,
\be
C(x_1,x_2) = \cases{{\displaystyle -{2 \over \pi x_{12}}} + \ldots,
\hbox{\ if $x_{12}$ is odd},\cr
\noalign{\medskip}
0, \hbox{\ if $x_{12}$ is even}.}
\ee
Again this characteristic difference in the parity of the distance between the two
monomers is expected, since on a finite rectangle, there must be an equal number of
monomers on the even sublattice as on the odd sublattice.

For alternating even and odd monomer positions $x_i$ (that is, all $x_i-x_{i+1}$ are odd),
the 4- and 6-point functions are equal to
\bea
&& C(x_1,\cdots,x_4) = {4 \over \pi^2} \Big\{{1 \over x_{12}x_{34}} +
{1 \over x_{14}x_{23}} \Big\} + \ldots\\
\noalign{\medskip}
&& C(x_1,\cdots,x_6) = \nonumber\\
&& \hspace{1.3cm} -{8 \over \pi^3} \Big\{{1 \over x_{12}x_{34}x_{56}} +
{1 \over x_{12}x_{36}x_{45}} + {1 \over x_{14}x_{23}x_{56}} -
{1 \over x_{14}x_{25}x_{36}} + {1 \over x_{16}x_{23}x_{45}} +
{1 \over x_{16}x_{25}x_{34}} \Big\} + \ldots \nonumber\\
\eea

In the scaling limit, these correlators exactly match those of a complex chiral free fermion $\psi$,
\be
\lim_{\rm scaling}  C(x_1,\cdots,x_{2n}) = \la \psi(x_1) \, \psi^\dagger(x_2) \, \psi(x_3) \, \psi^\dagger(x_4) \cdots \psi(x_{2n-1}) \, \psi^\dagger(x_{2n}) \ra,
\label{cft}
\ee
if the 2-point functions are normalized as
\bea
\la \psi(x) \psi^\dagger(y) \ra &=& \la \psi^\dagger(x) \psi(y) \ra = -{2 \over \pi (x-y)},\\
\la \psi(x) \psi(y) \ra &=& \la \psi^\dagger(x) \psi^\dagger(y) \ra = 0.
\eea

Physically the charged fermions $\psi \equiv \psi_{\rm e}$ and $\psi^\dagger \equiv \psi_{\rm o}$ can be interpreted respectively as the insertion of a monomer at an even position and at an odd position (or vice-versa), so that a globally neutral correlator indicates an equal number of even and odd monomers. Their real and imaginary parts, $\psi_1 = {1 \over \sqrt{2}}(\psi + \psi^\dagger)$ and $\psi_2 = {1 \over {\rm i}\sqrt{2}}(\psi - \psi^\dagger)$ can be viewed as two uncoupled Ising Majorana fermions which form a chiral conformal field theory with $c=1$.

We prove, in the next section, the equality (\ref{cft}) for an arbitrary value of $n$.


\section{Boundary monomer correlators}

We start with the cases where the correlators do not vanish identically: we place 
$2n$ monomers on the boundary, located at $(x_i,1)$, $n$ of which are at even (odd) positions. Without loss of generality, one may choose $x_1$ even, $x_2$ odd, $x_3$ even, and so on.

The matrix $B$ used in the previous section is not the most convenient choice to
carry out the general calculation. We slightly modify the entries of $B$ labelled by the monomer positions: we set the diagonal elements to 0 and connect the monomers by pairs, by setting $B_{(x_i,1),(x_{i+1},1)} = 1 = -B_{(x_{i+1},1),(x_i,1)}$ for all $i$ odd.
It means that the restriction of $K + B$ to the monomer sites is not the identity
matrix like in the previous section, but a direct sum of 2-by-2 blocks equal to
$\Big({0 \atop -1} \; {1 \atop 0} \Big)$, whose determinant remains equal to 1. The other off-diagonal elements of $B$ are as before.

This choice ensures that $B$ is also antisymmetric, and such that $B_{ij} = 0$ if $i-j$ has coordinates of equal parities. Since the matrix $K^{-1}$ has the same property, see (\ref{kinv}), it follows that $(\bi + K^{-1} B)_{ij} = 0$ if $i-j$
has coordinates of opposite parities. By an appropriate ordering of the site indices,
$\bi + K^{-1} B$ can thus be brought to a block-diagonal form.

The determinant to be computed has dimension $8n$ since $B$ has rank $8n$: there are $2n$ monomer sites, and each of them has three nearest neighbours. We label the sites as in Fig.2, using two types of roman indices, $a$ and $\underline a$, each type of label taking its values in $\{1,2,\cdots,4n\}$. The labelling is such that the differences $a-b$ or $\underline{a} - \underline{b}$ have coordinates of equal parities, and that differences $a-\underline b$ or $\underline{a} - b$ of unlike sites have coordinates of opposite parities. In this basis, the matrices $B$ and $K^{-1}$ are off-diagonal, $B^{}_{ab} = B^{}_{\underline{a}\underline{b}} = K^{-1}_{ab} = K^{-1}_{\underline{a}\underline{b}} = 0$, while the matrix $\bi + K^{-1}B$ is block-diagonal,
\be
(\bi + K^{-1}B)_{ij} = \pmatrix{(\bi + K^{-1}B)_{ab} & 0 \cr
0 & (\bi + K^{-1}B)_{\underline{a}\underline{b}}}.
\ee

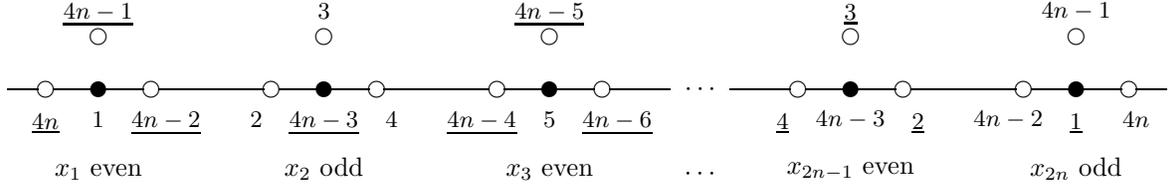
\begin{figure}[t]
\setlength{\unitlength}{1mm}
\begin{picture}(160,30)(-2,-15)
\put(0,0){\line(1,0){4}}
\put(6,0){\line(1,0){12}}
\put(20,0){\line(1,0){14}}
\put(36,0){\line(1,0){12}}
\put(50,0){\line(1,0){14}}
\put(66,0){\line(1,0){12}}
\put(80,0){\line(1,0){8}}
\put(90,0){\makebox(0,0)[l]{\small $\cdots$}}
\put(96,0){\line(1,0){8}}
\put(106,0){\line(1,0){12}}
\put(120,0){\line(1,0){14}}
\put(136,0){\line(1,0){12}}
\put(150,0){\line(1,0){4}}
\put(5,0){\circle{2}}
\put(12,0){\circle*{2}}
\put(19,0){\circle{2}}
\put(12,7){\circle{2}}
\put(5,-4.5){\makebox(0,0)[c]{\footnotesize {\underline{$4n$}}}}
\put(12,-4){\makebox(0,0)[c]{\footnotesize  1}}
\put(12,10){\makebox(0,0)[c]{\footnotesize {\underline{$4n-1$}}}}
\put(21,-4.5){\makebox(0,0)[c]{\footnotesize {\underline{$4n-2$}}}}
\put(12,-12){\makebox(0,0)[b]{\small $x_1$ even}}
\put(35,0){\circle{2}}
\put(42,0){\circle*{2}}
\put(49,0){\circle{2}}
\put(42,7){\circle{2}}
\put(33,-4){\makebox(0,0)[c]{\footnotesize {$2$}}}
\put(42,-4.5){\makebox(0,0)[c]{\footnotesize {\underline{$4n-3$}}}}
\put(42,10.5){\makebox(0,0)[c]{\footnotesize  3}}
\put(51,-4){\makebox(0,0)[c]{\footnotesize  4}}
\put(42,-12){\makebox(0,0)[b]{\small $x_2$ odd}}
\put(65,0){\circle{2}}
\put(72,0){\circle*{2}}
\put(79,0){\circle{2}}
\put(72,7){\circle{2}}
\put(63,-4.5){\makebox(0,0)[c]{\footnotesize {\underline{$4n-4$}}}}
\put(72,-4){\makebox(0,0)[c]{\footnotesize  5}}
\put(72,10){\makebox(0,0)[c]{\footnotesize {\underline{$4n-5$}}}}
\put(81,-4.5){\makebox(0,0)[c]{\footnotesize {\underline{$4n-6$}}}}
\put(72,-12){\makebox(0,0)[b]{\small $x_3$ even}}
\put(92,-12){\makebox(0,0)[b]{\small $\cdots$}}
\put(105,0){\circle{2}}
\put(112,0){\circle*{2}}
\put(119,0){\circle{2}}
\put(112,7){\circle{2}}
\put(103,-4.5){\makebox(0,0)[c]{\footnotesize {\underline{$4$}}}}
\put(112,-4){\makebox(0,0)[c]{\footnotesize {$4n-3$}}}
\put(112,10){\makebox(0,0)[c]{\footnotesize {\underline{$3$}}}}
\put(121,-4.5){\makebox(0,0)[c]{\footnotesize {\underline{$2$}}}}
\put(112,-12){\makebox(0,0)[b]{\small $x_{2n-1}$ even}}
\put(135,0){\circle{2}}
\put(142,0){\circle*{2}}
\put(149,0){\circle{2}}
\put(142,7){\circle{2}}
\put(133,-4){\makebox(0,0)[c]{\footnotesize {$4n-2$}}}
\put(142,-4.5){\makebox(0,0)[c]{\footnotesize {\underline{$1$}}}}
\put(142,10.5){\makebox(0,0)[c]{\footnotesize {$4n-1$}}}
\put(150,-4){\makebox(0,0)[c]{\footnotesize {$4n$}}}
\put(142,-12){\makebox(0,0)[b]{\small $x_{2n}$ odd}}
\end{picture}
\caption{Labelling of the $8n$ sites involved in the calculation of the determinant.
The solid circles represent the positions of the monomers ($x_1,x_2, \ldots$), all located on the boundary, the open circles their nearest neighbours.}
\end{figure}

Moreover the two diagonal blocks are closely related. As is manifest in Fig.2, the two types of sites are exchanged by a mirror symmetry, under which the monomer coordinates and the separation distances are transformed according to $x_\ell \to \tilde x_\ell = x_{2n+1-\ell}$ and $x_k - x_\ell \to \tilde x_\ell - \tilde x_k$. It follows that the second block depends on the separation distances $\tilde x_\ell - \tilde x_k$ in the same way the first block depends on $x_k-x_\ell$, or equivalently, 
\be
(\bi + K^{-1}(x_k-x_\ell)B)^{}_{\underline{a}\underline{b}} = (\bi + K^{-1}(\tilde x_\ell-\tilde x_k)B)^{}_{ab}\,.
\ee

As it turns out, the determinant of the first block will be invariant under the
substitution $x_k - x_\ell \to \tilde x_\ell - \tilde x_k$, and therefore equal to the
determinant of the second block. Putting all together, one obtains the correlations as
\be
C(x_1,x_2,\ldots,x_{2n}) = \sqrt{\det(\bi + K^{-1}B)_{ij}} = |\det(\bi + K^{-1}B)_{ab}|.
\ee
We finish the proof by showing that this last determinant reduces, in the
scaling limit, to the fermionic $2n$-point function (\ref{cft}).

We start by grouping the $4n$ sites $a$ by four, each group receiving a label
$I$, between 1 and $n$: $I=1$ corresponds to the first four sites $\{1,2,3,4\}$, $I=2$ to the next four sites $\{5,6,7,8\}$, and so on. Accordingly we write the matrix $(\bi + K^{-1}B)_{ab} = A_{IJ}$ in a block form, where all
blocks $A_{IJ}$ have dimension 4.

Because the matrix $B$ does not connect sites belonging to different values of
$I$, the diagonal blocks $A_{II}$ are all given in terms of a single matrix function
$A_1$, as are the off-diagonal blocks in terms of a second matrix $A_2$.
The entries of $A_{II}$ only depend on the distance $x_{2I-1,2I}$, so that $A_1$ depends on a single variable, $A_{II} = A_1(x_{2I-1,2I})$; likewise the off-diagonal block $A_{IJ}$ only depends on the two distances $x_{2I-1,2J}$ and $x_{2I,2J}$, so that $A_{IJ} = A_2(x_{2I-1,2J};x_{2I,2J})$ (to see that the other two distances $x_{2I-1,2J-1}$
and $x_{2I,2J-1}$ do not enter in $A_2$ requires a simple calculation, illustrated below).
Therefore the matrix of which we need to compute the determinant has the form
\be
(\bi + K^{-1}B)_{ab} = A_{IJ} = \pmatrix{
A_1(x_{12}) & A_2(x_{14};x_{24}) & A_2(x_{16};x_{26}) & \cdots \cr
A_2(x_{32};x_{42}) & A_1(x_{34}) & A_2(x_{36};x_{46}) & \cdots \cr
A_2(x_{52};x_{62}) & A_2(x_{54};x_{64}) & A_1(x_{56}) & \cdots \cr
\vdots & \vdots & \vdots & \ddots}.
\ee

Our next task is to compute the asymptotic value of the blocks $A_1$ and $A_2$, when the distances become large. This is straightforward as soon as one is familiar with the notations and with the form of $B$. Suppose that we want to compute the (1,1)-element of the off-diagonal block $A_{I=1,J=n}$ corresponding to the two subsets of sites fully displayed in Fig.2. The first labels in $I=1$ and $J=n$ are respectively the sites 1 and $4n-3$. One finds
\bea
(A_{I=1,J=n})_{(1,1)} \egal (\bi + K^{-1}B)_{1,4n-3} \nonumber\\
&& \hspace{-1cm} = K^{-1}_{1,\un 1}B^{}_{\un 1,4n-3} + K^{-1}_{1,\un 2}B^{}_{\un 2,4n-3} + K^{-1}_{1,\un 3}B^{}_{\un 3,4n-3} + K^{-1}_{1,\un 4}B^{}_{\un 4,4n-3} \nonumber\\
&& \hspace{-1cm} = - K^{-1}_{1,\un 1} - K^{-1}_{1,\un 2} - K^{-1}_{1,\un 3} + K^{-1}_{1,\un 4} \nonumber\\
&& \hspace{-1cm} = - K^{-1}_{(x_1,1),(x_{2n},1)} - K^{-1}_{(x_1,1),(x_{2n-1}+1,1)} - K^{-1}_{(x_1,1),(x_{2n-1},2)} + K^{-1}_{(x_1,1),(x_{2n-1}-1,1)}.
\eea
The last three terms cancel because they are equal to $(K^{-1}K)_{(x_1,1),(x_{2n-1},1)} = 0$, leaving, for $x \equiv x_1 - x_{2n}$,
\bea
(A_{I=1,J=n})_{11} \egal - K^{-1}_{(x_1,1),(x_{2n},1)} = -G^{\rm cl}(\textstyle{x-1 \over 2};1,1) + G^{\rm cl}({x+1 \over 2};1,1) \nonumber\\
\egal -G(\textstyle{x-1 \over 2},0) - G(\textstyle{x-1 \over 2},1) + G(\textstyle{x+1 \over 2},0) + G(\textstyle{x+1 \over 2},1) = -{2 \over \pi x} + \ldots
\eea
for large $x$ by using (\ref{expan}). Similar calculations for the other entries and for the diagonal blocks yield the matrices $A_1$ and $A_2$ explicitly as
\be
A_1(x) = \pmatrix{
-{2 \over \pi x} & {2 \over \pi x} & {2 \over \pi x} & -{2 \over \pi x}\cr
{1 \over \pi} & 1-{1 \over \pi} & -{1 \over \pi} & {1 \over \pi}\cr
1-{2 \over \pi} & -1+{2 \over \pi} & {2 \over \pi} & 1-{2 \over \pi}\cr
-{1 \over \pi} & {1 \over \pi} & {1 \over \pi} & 1-{1 \over \pi}} + \ldots, \quad
A_2(x;y) = \pmatrix{
-{2 \over \pi x} & {2 \over \pi x} & {2 \over \pi x} & -{2 \over \pi x}\cr
-{2 \over \pi y} & {2 \over \pi y} & {2 \over \pi y} & -{2 \over \pi y}\cr
0 & 0 & 0 & 0\cr
-{2 \over \pi y} & {2 \over \pi y} & {2 \over \pi y} & -{2 \over \pi y}} + \ldots
\ee
where the dots represent lower order terms in $x$ or $y$.

A first observation is that the full matrix $A_{IJ}$ contains exactly $n$ lines with all
their elements of order $-1$ in the distances $x_{\ell m}$, while all the other $3n$
lines contain elements of order 0 in these variables, coming from the $A_1$ blocks.
From this, it follows that the dominant term in the determinant has order $-n$ in the
distances. Anticipating that the coefficient of this term does not vanish so that the scaling dimension of the $2n$-correlator is $n$, we may neglect the $y$ dependence in the $A_2$-blocks, and use the simplified matrix,
\be
A_2(x) = A_2(x;\infty) = \pmatrix{
-{2 \over \pi x} & {2 \over \pi x} & {2 \over \pi x} & -{2 \over \pi x}\cr
0 & 0 & 0 & 0\cr 0 & 0 & 0 & 0\cr 0 & 0 & 0 & 0}.
\ee

At this stage, the full matrix has the following form,
\be
(\bi + K^{-1}B)_{ab} = A_{IJ} = \pmatrix{
A_1(x_{12}) & A_2(x_{14}) & A_2(x_{16}) & \cdots \cr
A_2(x_{32}) & A_1(x_{34}) & A_2(x_{36}) & \cdots \cr
A_2(x_{52}) & A_2(x_{54}) & A_1(x_{56}) & \cdots \cr
\vdots & \vdots & \vdots & \ddots}.
\ee
It has $n$ block columns, each formed of four columns. Within each block column, we uniformly add the first column to the second and third ones, and subtract it from the fourth one. This column operation does not change the value of the determinant, and can be done at the level of the small matrices $A_1$ and $A_2$. Doing this recasts the determinant into the following form,
\be
\det (\bi + K^{-1}B)_{ab} = \det \left(\begin{array}{cccc|cccc|cccc|cc}
-{2 \over \pi x_{12}} & 0 & 0 & 0 & -{2 \over \pi x_{14}} & 0 & 0 & 0 & -{2 \over
\pi x_{16}} & 0 & 0 & 0 & \ldots & \cr
{1 \over \pi} & 1 & 0 & 0 & 0 & 0 & 0 & 0 & 0 & 0 & 0 & 0 & \ldots & \cr
1 - {2 \over \pi} & 0 & 1 & 0 & 0 & 0 & 0 & 0 & 0 & 0 & 0 & 0 & \ldots & \cr
-{1 \over \pi} & 0 & 0 & 1 & 0 & 0 & 0 & 0 & 0 & 0 & 0 & 0 & \ldots & \cr
\hline
-{2 \over \pi x_{32}} & 0 & 0 & 0 & -{2 \over \pi x_{34}} & 0 & 0 & 0 & -{2 \over
\pi x_{36}} & 0 & 0 & 0 & \ldots \cr
0 & 0 & 0 & 0 & {1 \over \pi} & 1 & 0 & 0 & 0 & 0 & 0 & 0 & \ldots \cr
0 & 0 & 0 & 0 & 1 - {2 \over \pi} & 0 & 1 & 0 & 0 & 0 & 0 & 0 & \ldots \cr
0 & 0 & 0 & 0 & -{1 \over \pi} & 0 & 0 & 1 & 0 & 0 & 0 & 0 & \ldots \cr
\hline
-{2 \over \pi x_{52}} & 0 & 0 & 0 & -{2 \over \pi x_{54}} & 0 & 0 & 0 & -{2 \over
\pi x_{56}} & 0 & 0 & 0 & \ldots \cr
0 & 0 & 0 & 0 & 0 & 0 & 0 & 0 & {1 \over \pi} & 1 & 0 & 0 & \ldots \cr
0 & 0 & 0 & 0 & 0 & 0 & 0 & 0 & 1 - {2 \over \pi} & 0 & 1 & 0 & \ldots \cr
0 & 0 & 0 & 0 & 0 & 0 & 0 & 0 & -{1 \over \pi} & 0 & 0 & 1 & \ldots \cr
\hline
\vdots & \vdots & \vdots & \vdots & \vdots & \vdots & \vdots & \vdots & \vdots &
\vdots & \vdots & \vdots & \ddots
\end{array}\right).
\ee
It clearly factorizes into the product of two determinants. One has order $3n$, made of all rows and columns except the 1st, 5th, 9th, ..., and is equal to 1, while the other contains the remaining rows and columns. We therefore obtain
\be
\lim_{\rm scaling} C(x_1,\ldots,x_{2n}) = \lim_{\rm scaling} |\det (\bi + K^{-1}B)_{ab}| = \Big({-2 \over \pi}\Big)^n \; \det\Big({1 \over x_{2i-1} - x_{2j}}\Big)_{1 \leq i,j \leq n}\,.
\ee
This last form, a Cauchy determinant, can be evaluated explicitly (see for instance \cite{krat}),
\be
\lim_{\rm scaling} C(x_1,\ldots,x_{2n}) = \Big({-2 \over \pi}\Big)^n \; 
{\prod_{1 \leq i<j \leq n} \, (x_{2i-1}-x_{2j-1}) (x_{2j} - x_{2i}) \over \prod_{1 \leq i,j \leq n} \, (x_{2i-1} - x_{2j})},
\ee
and exactly reproduces the free fermion correlator (\ref{cft}).

We complete the proof by showing that all the correlators which do not contain an equal number of even and odd monomers vanish identically. Without loss of generality, one may assume that in addition to the $2n$ monomers at alternatively even and odd positions $(x_i,1)$, there are $M$ monomers, all located at either even or at odd positions $(y_i,1)$. We consider the case where the $M$ additional monomers are on even sites, the other case being similar. We assume $x_1 < x_2 < \cdots < x_{2n} < y_1 < \cdots < y_M$.

The matrix $(\bi + K^{-1}B)$ has now dimension $8n+4M$: the first $8n$ labels
will be as above, $4n$ indices $a$ and $4n$ indices $\un{a}$. The other $4M$
sites will be ordered in a more natural way: first monomer at $y_1$, its left,
upper and right neighbours, and so on for the others. The restriction of $B$
to the $8n$ sites is kept as above, with oriented bonds between $x$-monomers.
As $M$ can be odd, we take on the other $4M$ sites a direct sum of $B_1(y_i)$
matrices, written in (\ref{B1}).

Let us now compute the $n+M$ rows of $(\bi + K^{-1}B)$ labelled by the even $x$-monomers
(corresponding to $x_1,x_3,\ldots$) and by the $y$-monomers (in the case where all $y_i$
are odd, one would look instead at the odd $x$-monomers). A straightforward calculation shows that the rows labelled by the even monomers $(x_i,1)$, $i$ odd, are equal to
\bea
(\bi + K^{-1}B)_{(x_i,1),\cdot} &=& \Big(-K^{-1}_{(x_i,1),(x_2,1)},
K^{-1}_{(x_i,1),(x_2,1)}, K^{-1}_{(x_i,1),(x_2,1)}, -K^{-1}_{(x_i,1),
(x_2,1)}\;;\; \nonumber\\
&& \hspace{-2.5cm} -K^{-1}_{(x_i,1),(x_4,1)},K^{-1}_{(x_i,1),(x_4,1)},
K^{-1}_{(x_i,1),(x_4,1)}, -K^{-1}_{(x_i,1),(x_4,1)} \;;\; \cdots \;;\; \nonumber\\
&& \hspace{-2.5cm} -K^{-1}_{(x_i,1),(x_{2n},1)},K^{-1}_{(x_i,1),(x_{2n},1)},
K^{-1}_{(x_i,1),(x_{2n},1)}, -K^{-1}_{(x_i,1),(x_{2n},1)} \;;\; 0\,,\cdots,0
\;;\; 0\,,\cdots,0 \Big).
\label{xi}
\eea
The first $4n$ entries are non-zero, proportional to an matrix element of $K^{-1}$; then there is a group of $4n$ zeros corresponding to the underlined indices and another group
of zeros corresponding to the last $4M$ sites.

Because all $y_i$ are even, the first $8n$ entries of the rows $(\bi + K^{-1}B)_{y_i,\cdot}$ are exactly given by the formula (\ref{xi}) where one simply replaces $x_i$ by $y_i$. Moreover, one may check that the last $4M$ entries are also zero, so that one obtains the simple result that
\be
(\bi + K^{-1}B)_{(y_i,1),\cdot} = (\bi + K^{-1}B)_{(x_i,1),\cdot}\Big|_{x_i \to y_i}.
\ee

By column additions and subtractions, one may bring these rows to the form where
only $n$ entries are non-zero, for instance,
\bea
(\bi + K^{-1}B)_{(x_i,1),\cdot} &=& \Big(-K^{-1}_{(x_i,1),(x_2,1)},0,0,0\;;\;
-K^{-1}_{(x_i,1),(x_4,1)},0,0,0 \;;\; -K^{-1}_{(x_i,1),(x_{2n},1)},0,0,0 \;; \nonumber\\
&& \hspace{4cm} 0\,,\cdots,0 \;;\; 0\,,\cdots,0 \Big),
\eea
and similarly for $(\bi + K^{-1}B)_{(y_i,1),\cdot}$. In this form, the full matrix, which has the same determinant as the original one, contains $n+M$ rows which are vectors in an $\mathbb R^n$ vector subspace. Consequently, it has at least $M$ left eigenvectors with zero eigenvalue, therefore its determinant vanishes identically, and so does the corresponding correlator.


\section{String of monomers}

Up to here, we have considered isolated monomers, far apart from each other, and lying on the boundary of the upper-half plane. We have shown that the correlation functions for such monomer configurations can be understood, in the scaling limit, as free fermion correlators in a conformal field theory with central charge $c=1$. For completeness, we briefly discuss a different situation, and in some sense opposite, namely the case where the monomers form a compact cluster of consecutive sites. So we consider $2n$ monomers on consecutive boundary sites, and ask for the corresponding correlation $C_{2n}$, defined as before as the ratio of the partition function with the $2n$ monomers to the partition function with no monomer. 

Since the $2n$ monomers can be covered in a unique way by $n$ dimers, we may think of the dimer configurations in presence of the monomers as close-packed dimer configurations with a fixed string of $n$ consecutive dimers, all oriented along the boundary. Being now formulated as a pure dimer problem, with a prescribed boundary condition on an interval of length $2n$, the Temperley correspondence \cite{temp} (see also \cite{ipr}) with arrow configurations, or equivalently spanning trees, can be used. 

Given a dimer configuration, an associated configuration of arrows is defined on the odd-odd sublattice ${\cal L}_{\rm odd}$ (it contains a half of the boundary sites of the upper-half plane): if a dimer touches a site of ${\cal L}_{\rm odd}$, one draws an arrow from that site towards its nearest neighbour in ${\cal L}_{\rm odd}$, in the direction of the dimer; those dimers which do not touch sites of ${\cal L}_{\rm odd}$ are uniquely fixed once the dimers which do touch ${\cal L}_{\rm odd}$ are given. The so-obtained arrow configuration has the property that it cannot form closed loops, because a loop would encircle an odd number of sites of the original lattice, which therefore could not be fully covered by dimers. So the arrow configuration defines a spanning tree on ${\cal L}_{\rm odd}$.

In this correspondence, a prescribed string of $n$ consecutive dimers on the boundary translates into a string of $n$ consecutive arrows, all pointing to the left or to the right along the boundary. Elsewhere on the boundary, arrows are free to point in any of the three available directions (with the only constraint that the full arrow configuration cannot contain loops). This problem of arrows with a fixed string of aligned arrows on the boundary has been recently examined in \cite{wind} in the context of the Abelian sandpile model. It has been found that the correlation function $C_{2n}$ behaves asymptotically as
\be
C_{2n} \simeq A \: n^{-1/4} \: {\rm e}^{-2n{\rm G}/\pi}, \qquad n \: \rm large,
\label{c2n}
\ee
where ${\rm G} = 0.915965$ is the Catalan constant, and $A$ is a numerical constant. The exponential decay is expected and due to the defect of entropy of the $n$ boundary sites which have their arrow frozen. On the other hand, the power law $n^{-1/4}$ has been understood within conformal field theory, as the correlator of two boundary condition changing fields (and an extra dimension 0 field corresponding to the insertion of dissipation, see \cite{piru}). One of these two fields changes the free arrow boundary condition into say, the right arrow boundary condition, and has dimension  $-1/8$, the other changes the right arrow boundary condition back into the free arrow boundary condition, and has dimension $3/8$. The two dimensions add up to $1/4$, and account for the exponent in (\ref{c2n}).

Note that these dimensions have been obtained in \cite{wind} in the context of the sandpile model (equivalently spanning trees), known to correspond to a (logarithmic) conformal field theory with central charge $c=-2$. However, as discussed in the introduction, various results, including those of the present article, point to a description of the general monomer-dimer problem in terms of a conformal field theory with central charge $c=1$, of which the theory with $c=-2$ would appear as a subtheory accounting for the dimer degrees of freedom. In the $c=1$ setting, the above power law is presumably interpreted as a 2-point function of fields with dimension $1/8$.


\acknowledgments
This work was supported by a RFBR grant No 06-01-00191a, and by the Belgian Interuniversity Attraction Poles Program P6/02, through the network NOSY (Nonlinear systems, stochastic processes and statistical mechanics). V.B.P. would like to thank the Universit\'e catholique de Louvain for its financial support and hospitality. P.R. is a Research Associate of the Belgian National Fund for Scientific Research (FNRS).



\begin{thebibliography}{99}

\bibitem{foru}  R.H. Fowler and G.S. Rushbrooke, Trans. Faraday Soc. {\bf 33}, 1272
(1937).

\bibitem{kast}  P.W. Kasteleyn, Physica {\bf 27}, 1209 (1961); J. Math. Phys.
{\bf 4}, 287 (1963).

\bibitem{fish} M.E. Fisher, Phys. Rev. {\bf 124}, 1664 (1961).

\bibitem{fite} H.N.V. Temperley and M.E. Fisher, Phil. Mag. {\bf 6}, 1061 (1961).

\bibitem{ferd}  A.E. Ferdinand, J. Math. Phys. {\bf 8}, 2332 (1967).

\bibitem{wu1} F.Y. Wu, Phys. Rev. {\bf 168}, 539 (1968).

\bibitem{fist} M.E. Fisher and J. Stephenson, Phys. Rev. {\bf 132}, 1411 (1963); see also
R.E. Hartwig, J. Math. Phys. {\bf 7}, 286 (1966).

\bibitem{tzwu} W.J. Tseng and F.Y. Wu, J. Stat. Phys. {\bf 110}, 671 (2003).

\bibitem{kong1} Y. Kong, Phys. Rev. E {\bf 74}, 011102 (2006).

\bibitem{kong2} Y. Kong, {\sl Logarithmic corrections in the free energy of monomer-dimer model on plane lattices with free boundaries}, arXiv:cond-mat/0601568.

\bibitem{kong3} Y. Kong, {\sl Monomer-dimer model in two-dimensional rectangular lattices with fixed dimer density}, arXiv:cond-mat/0610690.

\bibitem{fend} P. Fendley, R. Moessner and S.L. Sondhi, Phys. Rev. B {\bf 66}, 214513
(2002).

\bibitem{FanWu} Ch. Fan and F.Y. Wu, Phys.Rev. B {\bf 2}, 723 (1970).

\bibitem{iprh} N.Sh. Izmailian, V.B. Priezzhev, P. Ruelle and C.-K. Hu, Phys. Rev.
Lett. {\bf 95}, 260602 (2005).



\bibitem{auperk} H. Au-Yang and J.H.H. Perk, Phys. Lett. A {\bf 104}, 131 (1984).

\bibitem{ken} R. Kenyon, Ann. Probab. {\bf 29}, 1128 (2001).

\bibitem{dijk} R. Dijkgraaf, D. Orlando and S. Reffert, {\sl Dimer models, free fermions and super quantum mechanics}, arXiv:0705.1645 [hep-th].

\bibitem{krat} C. Krattenthaler, S\'em. Lothar. Combin. {\bf 42}, B42q (1999).

\bibitem{temp} H.N.V. Temperley, in {\it Combinatorics: Proceedings of the British
Combinatorial Conference}, London Math. Soc. Lecture Notes Series 13, 1974, p. 202.

\bibitem{ipr} N.Sh. Izmailian, V.B. Priezzhev and P. Ruelle, SIGMA {\bf 3}, 001 (2007).

\bibitem{wind} P. Ruelle, J. Stat. Mech. P09013 (2007).

\bibitem{piru} G. Piroux and P. Ruelle, J. Stat Mech. P10005 (2004).

\end{thebibliography}
\end{document}